\documentclass[aps,prl,floatfix,showpacs,twocolumn]{revtex4-1}
\usepackage{dcolumn}
\usepackage{bm}
\usepackage{color}
\usepackage{graphicx}
\usepackage{epsf}
\usepackage{pstricks}
\usepackage{amsmath}
\usepackage{hyperref}
\begin{document}
\title{Electromagnetic and neutral-weak response functions of $^{4}$He and $^{12}$C}
\author{
A.\ Lovato$^{\, {\rm a,b} }$,
S.\ Gandolfi$^{\, {\rm c} }$,
J.\ Carlson$^{\, {\rm c} }$,
Steven\ C.\ Pieper$^{\, {\rm b} }$,
and R.\ Schiavilla$^{\, {\rm d,e} }$
}
\affiliation{
$^{\,{\rm a}}$\mbox{Argonne Leadership Computing Facility, Argonne National Laboratory, Argonne, IL 60439}\\
$^{\,{\rm b}}$\mbox{Physics Division, Argonne National Laboratory, Argonne, IL 60439}\\
$^{\,{\rm c}}$\mbox{Theoretical Division, Los Alamos National Laboratory, Los Alamos, NM 87545}\\
$^{\,{\rm d}}$\mbox{Theory Center, Jefferson Lab, Newport News, VA 23606}\\
$^{\,{\rm e}}$\mbox{Department of Physics, Old Dominion University, Norfolk, VA 23529}
}
\date{\today}

\begin{abstract}
{\it Ab initio} calculations of the quasi-elastic electromagnetic and neutral-weak
response functions of $^4$He and $^{12}$C are carried out for the first time.
They are based on a realistic approach to nuclear dynamics, in which the strong
interactions are described by two- and three-nucleon potentials and the electroweak
interactions with external fields include one- and two-body terms. The Green's function
Monte Carlo method is used to calculate directly the Laplace transforms of the response
functions, and maximum-entropy techniques are employed to invert the resulting
imaginary-time correlation functions with associated statistical errors. The theoretical 
results, confirmed by experiment in the electromagnetic case, show that two-body currents 
generate excess transverse strength from threshold to the quasi-elastic to the dip region 
and beyond. These findings challenge the conventional picture of quasi-elastic inclusive
scattering as being largely dominated by single-nucleon knockout processes.
\end{abstract} 

\pacs{21.60.De, 25.30.Pt}

\index{}\maketitle

In first-order perturbation theory, the interactions
of an external electroweak probe with a nucleus are described
by response functions.  These response functions---two
for the processes $A(e,e^\prime)$ induced by electromagnetic interactions,
and five for the processes $A(\nu_l,\nu_l^\prime)$ and $A(\overline{\nu}_l,\overline{\nu}_l^{\, \prime})$,
or $A(\nu_l,l^-)$ and $A(\overline{\nu}_l,l^+)$, induced by neutral or
charge-changing weak interactions---determine the inclusive
differential cross sections~\cite{Shen:2012}.  They can be written
schematically as
\begin{equation}
R_{\alpha\beta}(q,\omega)\!\sim\! \sum_f \delta(\omega+E_0-E_f)
\langle f \vert O_\alpha({\bf q}) \vert 0\rangle^* \langle f \vert O_\beta({\bf q}) \vert 0\rangle ,
\label{eq:respon}
\end{equation}
where ${\bf q}$ and $\omega$ are the momentum and energy transfers
injected by the external field into the nucleus, $\vert 0\rangle$ and $\vert f\rangle$
represent respectively its initial ground state of energy $E_0$ and
final continuum state of energy $E_f$, $O_{\alpha}({\bf q})$ and $O_{\beta}({\bf q})$ denote
appropriate components of the of the nuclear electroweak current operator
(their $\omega$-dependence is dealt with as described below), and
an average over the ground-state spin projections is understood (precise
definitions for the nuclear electroweak response functions, and resulting
inclusive cross sections, are given in Ref.~\cite{Shen:2012}).

At large values of momentum and energy transfers ($q\gtrsim 1$ GeV
and $\omega \gtrsim 0.5$ GeV), where the dynamics of interacting nucleons 
is inextricably interwoven with the internal dynamics of individual nucleons,
the accurate calculation of the response functions poses formidable
challenges, particularly in view of the fact that a consistent theoretical
framework to describe such a regime is still lacking.  Even at the lower $q$
and $\omega$ of interest in the present study ($q \lesssim 0.5$ GeV and
$\omega$ in the quasi-elastic region), where the consequences of the nucleon's
substructure on nuclear dynamics can be subsumed into effective many-body
potentials and currents, this calculation remains extremely difficult:
it requires knowledge of the whole excitation spectrum of the nucleus
and inclusion in the electroweak currents of one- and many-body terms. 

In the case of inclusive weak scattering, a further difficulty exists
for comparing calculated and experimental results.  The experimental
initial neutrino energy is not known; instead there is a broad spectrum
of energies.  This means that
the observed cross section for a given energy and angle of the final
lepton follows from a folding with the energy distribution of the incoming
neutrino flux and, consequently, may include contributions from $q$-$\omega$
regions where different mechanisms are at play: the threshold region, where 
the structure of the low-lying energy spectrum and collective effects
are important; the quasi-elastic region, which is dominated by scattering
off individual nucleons and nucleon pairs (see below); and the $\Delta$-resonance
region, where one or more pions are produced in the final state~\cite{Benhar:2012}.

Integral properties of the response functions can be studied by
means of sum rules, which are obtained from ground-state
expectation values of appropriate combinations of the current operators
(and commutators of these combinations with the Hamiltonian
in the case of energy-weighted sum rules), thus avoiding the need
for computing the nuclear excitation spectrum.  {\it Ab initio} quantum Monte Carlo (QMC)
calculations of (non-energy-weighted) electroweak sum rules in $^{12}$C
have been recently reported in Refs.~\cite{Lovato:2013,Lovato:2013b}.
These calculations have demonstrated that a large fraction ($\simeq 30$\%)
of the strength in the response arises from processes involving two-body
currents, and that interference effects between the matrix elements of
one- and two-body currents play a major role \cite{Benhar:2013}.
These effects are typically only partially, or approximately, accounted for in 
existing perturbative or mean-field studies~\cite{Martini:2009,
Martini:2010,Nieves:2011,Amaro:2011}.

Yet, sum rules do not provide direct information on the distribution of strength,
whether, for example, the calculated excess strength induced by two-body
currents is mostly at large $\omega$, well beyond the quasi-elastic peak energy
$\omega_{\rm qe}=\sqrt{q^2+m^2}-m$ ($m$ is the nucleon mass), or is also found
in the quasi-elastic region with $\omega \lesssim \omega_{\rm qe}$.  Moreover,
in the electromagnetic case, comparison of theoretical and experimental sum rules is
problematic, since longitudinal and transverse response functions obtained from
Rosenbluth separation of measured inclusive $(e,e^\prime)$ cross sections are only
available in the space-like region ($\omega < q$) and therefore must be extrapolated
into the unobserved time-like region ($\omega > q$) before ``experimental''
values for the sum rules can be determined, see Refs.~\cite{Lovato:2013,Carlson:2002}
for a discussion of these issues.

In this paper we report the first {\it ab initio} calculations
of the electromagnetic and neutral-weak response functions of
$^4$He and $^{12}$C (other studies for $^4$He have been already performed
within different frameworks, i.e.~\cite{Leidemann:2013,Bacca:2009}).
These calculations proceed in two steps:
the first involves the use of QMC methods to compute the response
in imaginary time, the so-called Euclidean response~\cite{Carlson:1992,Carlson:1994},
while the second consists in the inversion, via maximum entropy
techniques~\cite{Bryan:1990,Jarrell:1996}, of these ``noisy'' imaginary-time
data to obtain $R_{\alpha\beta}(q,\omega)$.  The dynamical framework
is based on a realistic Hamiltonian, including the Argonne $v_{18}$
two-nucleon~\cite{Wiringa:1995} (AV18) and Illinois-7 three-nucleon~\cite{Pieper:2008}
(IL7) potentials, and on realistic electroweak currents with one- and
two-body terms.  A concise description of this framework is in Refs.~\cite{Lovato:2013,Lovato:2013b},
while a more extended one can be found in the reviews~\cite{Carlson:1998,Carlson:2014}.
These latter papers also illustrate the level of quantitative success it has achieved
in accurately predicting many properties of s- and p-shell nuclei
up to $^{12}$C, including, among others, energy spectra of low-lying
states, static properties like charge radii, magnetic dipole and electric quadrupole
moments, radiative and weak transition rates, and elastic and inelastic
electromagnetic form factors.

The Euclidean response function is defined as the Laplace transform 
\begin{equation}
E_{\alpha\beta}(q,\tau) = C_{\alpha\beta}(q)\int_{\omega_{\rm th}}^\infty d\omega\,
 e^{-\tau \omega} R_{\alpha\beta}(q,\omega) \ ,
\label{eq:laplace_def}
\end{equation}
where $\omega_{\rm th}$ is the inelastic threshold and the $C_{\alpha\beta}$ are
$q$-dependent normalization factors.   In $R_{\alpha\beta}(q,\omega)$ the
$\omega$-dependence enters via the energy-conserving $\delta$-function and
the dependence on the four-momentum transfer $Q^2=q^2-\omega^2$ of the
electroweak form factors of the nucleon and $N$-to-$\Delta$ transition in the
currents.  We remove the latter by evaluating these form factors at
$Q^2_{\rm qe}=q^2-\omega_{\rm qe}^2$.  In the case of the electromagnetic
longitudinal ($L$ or $\alpha\beta=00$) and transverse ($T$ or $\alpha\beta=xx$)
response functions, the normalization factors are~\cite{Carlson:2002}
$C_L=C_T= 1/\left[G_{E}^{p}(Q^2_{\rm qe})\right]^2$, where $G_E^p$ is the
proton electric form factor, while in the neutral-weak response functions
they are the same as those adopted in the sum rule calculations reported in
Ref.~\cite{Lovato:2013b}.  With these definitions the response functions in
Eq.~(\ref{eq:laplace_def}) can be thought of as being due to
point-like, but strongly interacting, nucleons.  Note that non-energy-weighted
sum rules~\cite{Lovato:2013,Lovato:2013b} correspond to $E_{\alpha\beta}(q,\tau\!=\!0)$,
while energy-weighted ones are obtained by taking derivatives of $E_{\alpha\beta}(q,\tau)$
with respect to $\tau$ and evaluating them at $\tau=0$.

The Euclidean response can be expressed as a ground-state expectation
value, 
\begin{equation}
\frac{E_{\alpha\beta}(q,\tau)}{C_{\alpha\beta}(q)}= \frac{\langle 0| O^\dagger_{\alpha}({\bf q}) e^{-(H-E_0)\tau} 
O_{\beta}({\bf q}) |0\rangle}{\langle 0| e^{-(H-E_0)\tau}|0\rangle}\ ,
\label{eq:euc_me}
\end{equation}
where $H$ is the nuclear Hamiltonian (here, the AV18/IL7 model), $\tau$ is the imaginary-time, 
and $E_0$ is a trial energy to control the normalization.
In this paper we report responses computed with the variational wave function,
$\vert 0\rangle = \vert\Psi_V\rangle$; in Refs.~\cite{Lovato:2013,Lovato:2013b} it was
shown that sum rules computed with $\vert\Psi_V\rangle$ are very close
to those computed with the exact Green's function Monte Carlo (GFMC) wave functions.
The calculation of the
matrix element above is carried out with GFMC methods~\cite{Carlson:1992}
similar to those used in projecting out the exact ground state of $H$ from a trial
state~\cite{Carlson:1987}.  It proceeds in two steps.  First, an unconstrained imaginary-time
propagation of the variational Monte Carlo (VMC) state $\vert \Psi_V\rangle$ is performed
and saved. Next, the states $O_{\beta}({\bf q}) \vert\Psi_V\rangle$ are evolved in imaginary
time following the path previously saved.  During this latter imaginary-time evolution, scalar
products of ${\rm exp}\left[-\left(H-E_0\right) \tau_i\right]O_{\beta}({\bf q}) \vert\Psi_V\rangle$
with $O_{\alpha}({\bf q})\vert \Psi_V \rangle$ are evaluated on a grid of $\tau_i$ values, and
from these scalar products estimates for $E_{\alpha\beta}(q,\tau_i)$ are
obtained (a complete discussion of the methods is in Refs.~\cite{Carlson:1992,Carlson:2002}).
\begin{center}
\begin{figure}[bth!]
\includegraphics[width=9cm]{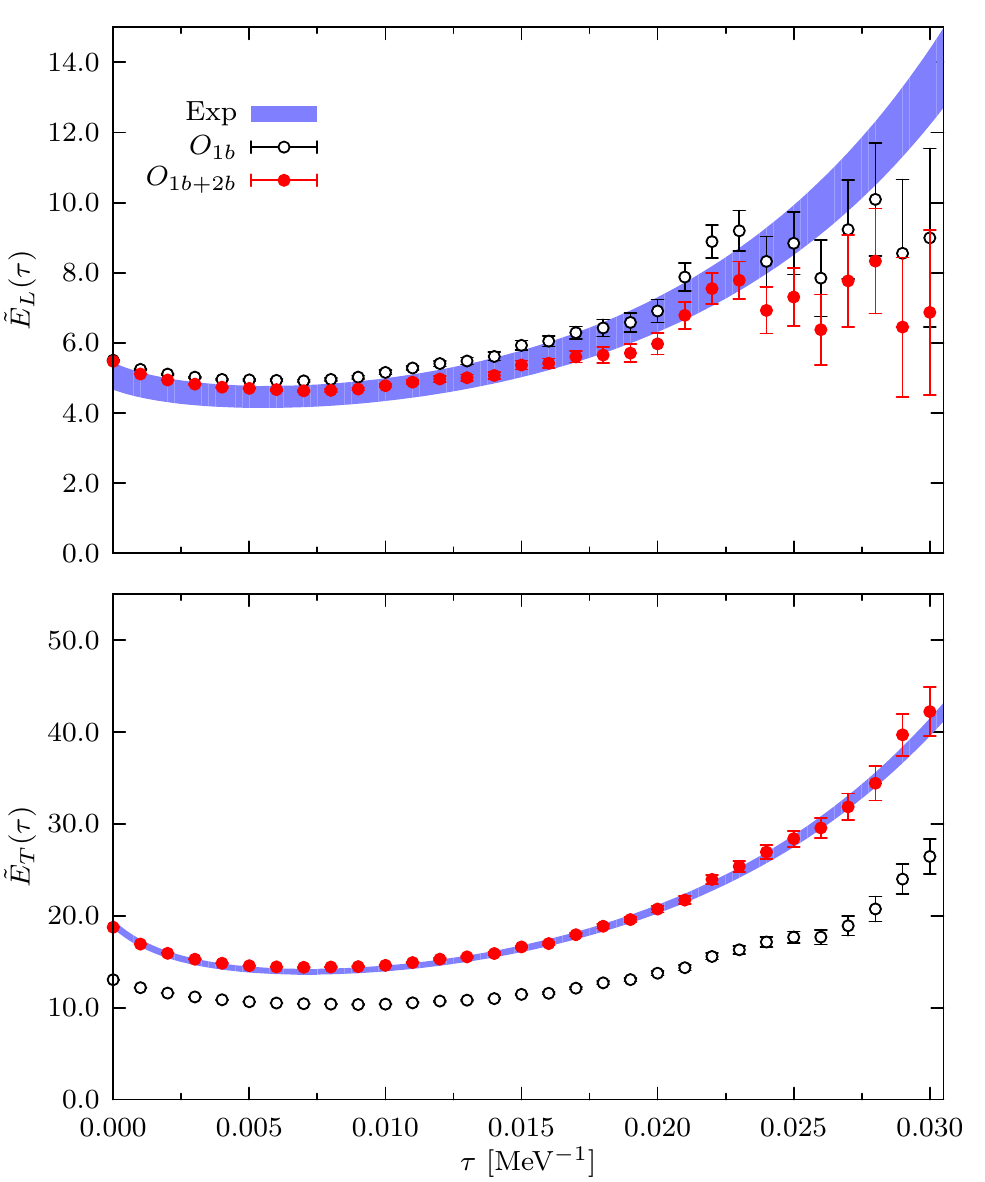}
\caption{(Color online) Euclidean electromagnetic longitudinal (top panel) and transverse (lower panel)
response function of $^{12}$C at $q=570$ MeV.  Experimental data are from Ref.~\cite{Jourdan:1996}.}
\label{fig:f1}
\end{figure}
\end{center}

In Fig.~\ref{fig:f1} the electromagnetic longitudinal ($E_L$, top panel) and 
transverse ($E_T$, lower panel) Euclidean response functions of $^{12}$C are compared to
those extracted from the world data analysis by Jourdan~\cite{Jourdan:1996},
represented by the shaded bands. In order to better show the
large $\tau$ behavior, all the figures in this paper show
$\widetilde{E}_{\alpha\beta}(q,\tau) = \exp[\tau\, q^2/(2m)] E_{\alpha\beta}(q,\tau)$; 
this scaled response would be a constant for an isolated proton.
The ``experimental'' $E_L(q,\tau)$ and
$E_T(q,\tau)$ follow from Laplace-transforming the longitudinal and transverse
data.  These are first divided by $\left[G_{E}^{p}(Q^2)\right]^2$ to obtain
corresponding response functions of point-like nucleons, and then integrated
with the weight factor ${\rm exp}(-\tau \omega)$ up to $\omega_{\rm max}$,
where measurements are available.  The strength in the unobserved region with
$\omega > \omega_{\rm max}$ is estimated by assuming that the
$R_L(q,\omega > \omega_{\rm max})$ and $R_T(q,\omega > \omega_{\rm max})$
of $^{12}$C are proportional to those in the deuteron, which can be accurately
calculated~\cite{Shen:2012}.  The procedure is identical to that used in
Ref.~\cite{Lovato:2013} for the sum rules.  As discussed in Ref.~\cite{Lovato:2013},
the scaling assumption can be justified by observing that the high $\omega$
(well beyond $\omega_{\rm qe}$) region of the response is dominated by
two-nucleon physics, in particular by deuteron-like $np$ pairs in the ground-state
of the nucleus.   It is important to stress that, as $\tau$ increases, the Euclidean
response functions become more and more sensitive to strength in the quasi-elastic
and threshold regions of $R_{L,T}(q,\omega)$.  Indeed, in this limit
($\tau \gtrsim 1/\omega_{\rm qe}$) contributions from unmeasured
strength at $\omega > \omega_{\rm max}$ are exponentially suppressed.

In Fig.~\ref{fig:f1} we show results obtained
by including only one-body (open circles) or both one- and two-body (solid circles)
terms in the electromagnetic transition operators.  In the longitudinal case, destructive
interference between the matrix elements of the one- and two-body charge operators
reduces, albeit slightly, the one-body response.  In the transverse case, on the other
hand, two-body current contributions substantially increase the one-body response.
This enhancement is effective over the whole imaginary-time region we have
considered, with the implication that excess transverse strength is generated by
two-body currents not only at $\omega \gtrsim \omega_{\rm qe}$, but also in the
quasi-elastic and threshold regions of $R_T(q,\omega)$.  It is reassuring to see
that the full predictions for both longitudinal and transverse Euclidean response
functions are in excellent agreement with data.

At larger values of $\tau$ the  statistical errors associated with the GFMC evolution
are rather large, particularly in the longitudinal response for which the
elastic contribution proportional to the square of the $^{12}$C form factor~\cite{Lovato:2013} needs
to be removed in order to account for the inelastic strength only.  However, it
should be possible to reduce these errors in the future by investing
substantial additional computational resources in this type of calculation.
%
% Blue could be removed, and remaining para joined with previous one
%
Those presented here were performed with $\sim$45 million core hours of 
Argonne National Laboratory's IBM Blue Gene/Q (Mira)
parallel supercomputer.  The Automatic Dynamic Load Balancing (ADLB) library~\cite{Lusk:2010}
was used to distribute the imaginary time propagation of $O_{\beta}(\mathbf{q})
\vert\Psi_V\rangle$ and the evaluation of the matrix element in Eq.~(\ref{eq:euc_me}) 
over more than 8000 MPI ranks. The code is at present approximately 75\% 
efficient at this scale.

\begin{center}
\begin{figure}[bth]
\includegraphics[width=8.5cm]{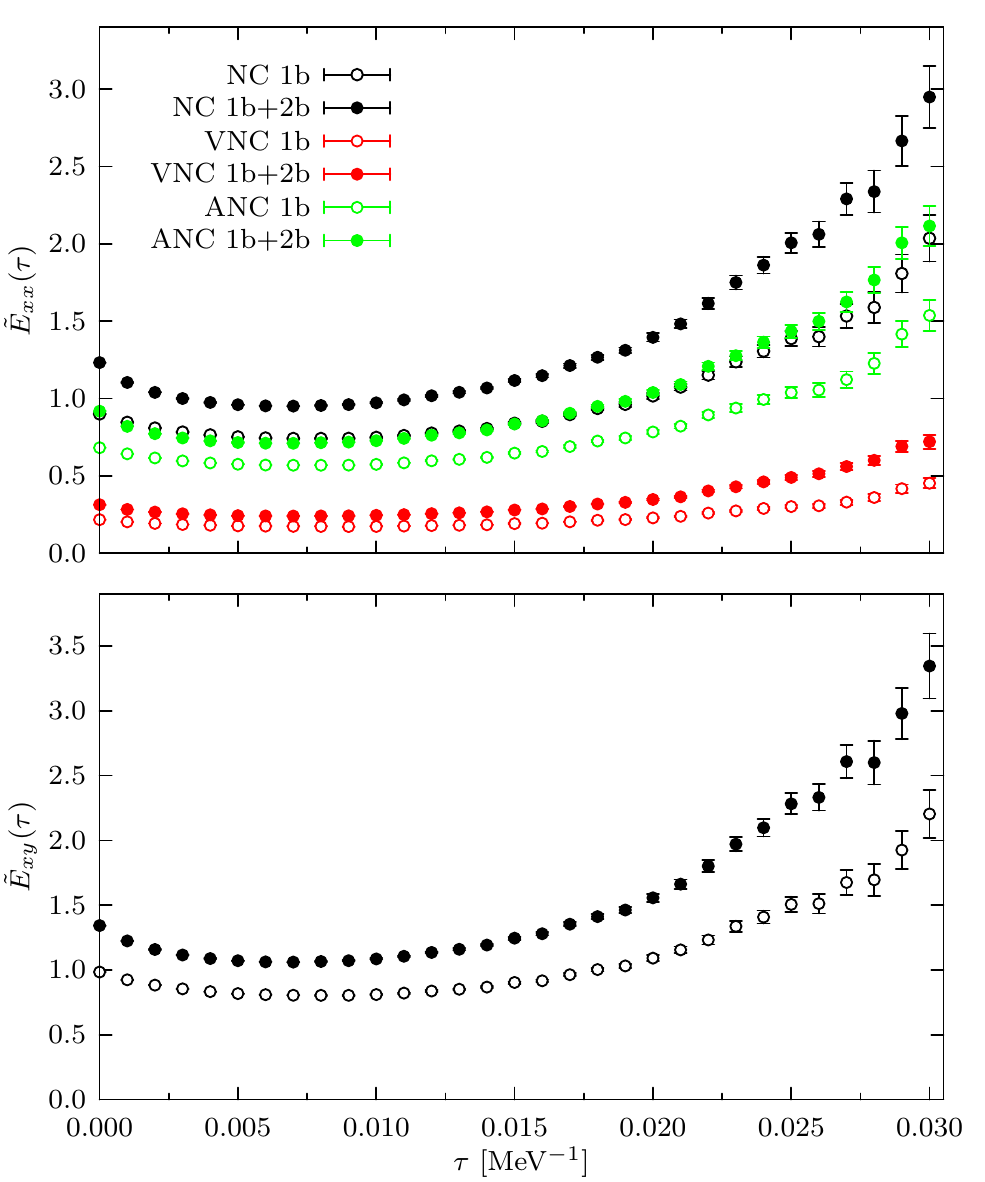}
\caption{(Color online) Euclidean neutral-weak transverse (top panel) and interference (lower panel)
response functions ($\alpha\beta =xx$ and $xy$ in the notation of Ref.~\cite{Shen:2012}) of $^{12}$C
at $q=570$ MeV.  See text for further explanations.}
\label{fig:f3}
\end{figure}
\end{center}

In Fig.~\ref{fig:f3} we show the largest of the five 
Euclidean neutral-weak response functions: the transverse (top panel) and interference (lower panel)
$E_{\alpha\beta}(q,\tau)$, having respectively $\alpha\beta=xx$ and
$\alpha\beta=xy$ in the notation of Ref.~\cite{Shen:2012}.  The $E_{xy}(q,\tau)$
response is due to interference between the vector (VNC) and axial (ANC)
parts of the neutral current (NC), and in the inclusive cross section the 
corresponding $R_{xy}(q,\omega)$ enters with opposite sign depending
on whether the process $A(\nu_l,\nu^\prime_l)$ or $A(\overline{\nu}_l,\overline{\nu}_l^{\,\prime})$
is considered~\cite{Shen:2012}.  On the other hand, in the transverse case
the interference of VNC and ANC terms vanishes, and $E_{xx}(q,\tau)$ is simply given
by the sum of the terms with both $O_\alpha$ and $O_\beta$ in Eq.~(\ref{eq:respon}) being
from the VNC or from the ANC.
For $E_{xx}(q,\tau)$ these individual contributions, along with their sum, are displayed 
separately.  Both $E_{xx}(q,\tau)$ and $E_{xy}(q,\tau)$ response functions
obtained with one-body terms only in the NC are substantially increased when
two-body terms are also retained.  This enhancement is found not only at low
$\tau$, thus corroborating the sum-rule predictions of Ref.~\cite{Lovato:2013b},
but in fact extends over the whole $\tau$ region studied here.  Moreover, in the case of
the transverse response it affects, in relative terms, the individual (VNC-VNC)
and (ANC-ANC) contributions about equally.
The VNC consists of a linear combination of the isoscalar and isovector
components of the electromagnetic current, weighted
respectively by the factors $-2\,\sin^2\theta_W$ and $(1 -2\,\sin^2\theta_W)$
with $\theta_W$ being the Weinberg angle.  The excess transverse strength
induced by two-body terms in the VNC is consistent with that found in the
transverse electromagnetic response, and is confirmed by experiment as
Fig.~\ref{fig:f1} demonstrates.  The two-body enhancement
in the (ANC-ANC) contribution of $E_{xx}(q,\tau)$ is substantial at these
relatively large $q$'s.  It decreases significantly (for $\tau \gtrsim 0.01$ MeV$^{-1}$)
as $q$ is reduced~\cite{Lovato:2015},
consistently with what is found in calculations of low $q$ charge-changing
weak transitions to specific low-lying states, such as the $\beta$-decays and
electron and muon captures studied in Refs.~\cite{Schiavilla:2002,Marcucci:2011},
where it amounts to a few percent.  In principle, the
enhancement in the quasi-elastic region could be measured
in parity-violating inclusive $(\vec{e}, e^\prime)$ scattering at
backward angles.  However, the smallness of the factor $(1 -4\,\sin^2\theta_W)$,
to which the relevant (VEM-ANC) interference response function is proportional,
makes experiments of this type extremely difficult.

In order to obtain more detailed information on the energy
dependence of the $R_{\alpha\beta}(q,\omega)$
response, we employ the maximum entropy (MaxEnt) method to invert
$E_{\alpha\beta}(q,\tau)$.
We describe the method here very briefly, several standard references are 
available~\cite{Bryan:1990,Jarrell:1996}.
The numerical inversion of a Laplace transform $E_{\alpha\beta}(q,\tau)$
with its associated statistical errors is a notoriously ill-posed problem.  
The fact that we are interested in the (smooth) response around the
quasi-elastic peak rather than isolated peaks makes it somewhat more
practical.  The MaxEnt method is based
on Bayesian statistical inference:
the ``most probable'' response function is the one that maximizes
the {\it posterior probability} Pr$[R\vert\overline{E}\, ]$, i.e., the conditional probability of $R$
given $\overline{E}$.  Bayes theorem states that the posterior probability is proportional to the
product ${\rm Pr}[\,\overline{E}\vert R\,]\times {\rm Pr}[R\,]$, where Pr$[\,\overline{E}\vert R\,]$
is the {\it likelihood function} and Pr$[R\,]$ is the {\it prior probability}.  Arguments based on the
central limit theorem show that the asymptotic limit of the likelihood function is given by
Pr$[\,\overline{E}\vert R\,] \propto {\rm exp}(-\chi^2/2)$ with $\chi^2$ defined as follows.
Let $N_\tau$ and $N_\omega$ be the numbers of grid points in the variables $\tau$ and
$\omega$, respectively.  Then the Laplace transform in Eq.~(\ref{eq:laplace_def}) reads
(the $q$-dependence and subscripts $\alpha\beta$ of $E_{\alpha\beta}(q,\tau)$ and
$R_{\alpha\beta}(q,\tau)$ are suppressed for simplicity hereafter)
\begin{equation}
E_i=\sum_{j=1}^{N_\omega} K_{ij}\, R_j \ ,
\label{eq:laplace_discrete}
\end{equation}
where $K_{ij}={\rm exp}(-\tau_i \, \omega_j)$ and $R_j=  \Delta\omega_j \, R(\omega_j)$,
and the $\chi^2$ follows from
\begin{equation}
\chi^2=\sum_{i,j=1}^{N_\tau} \left(\, \overline{E}_i-E_i\right)
\left(C^{-1}\right)_{ij} \left(\, \overline{E}_j-E_j\right)\ ,
\label{eq:chi2}
\end{equation}
where the $E_i$ are obtained from Eq.~(\ref{eq:laplace_discrete}),
the $\overline{E}_i$ are the GFMC calculated values, and $C$ is the
covariance matrix.  Therefore, maximizing the likelihood function
reduces to finding a set of $R_i$ values that minimizes the $\chi^2$.
The GFMC errors on $\overline{E}_i$ are strongly correlated in $\tau$, as
individual steps involve only small spatial distances and evolutions
of the spin-isospin amplitudes. It is therefore of
paramount importance to estimate the covariance matrix $C$.

Limiting ourselves only to the $\chi^2$ minimization would
implicitly be making the assumption that the prior probability is either
unimportant or unknown.  However, since the response function is positive
definite and normalizable, it can be interpreted as yet another probability 
function. The  {\it principle of maximum entropy} states that the values of a probability 
function are to be  assigned by maximizing the entropy
\begin{equation}
S = \sum_{i=1}^{N_\omega} \Big[ 
R(\omega_i)-M(\omega_i)-R(\omega_i)\ln[R(\omega_i)/M(\omega_i)] \Big] \Delta\omega_i\, ,
\end{equation}
where the positive definite function $M(\omega)$ is the {\it default model}.  It is worthwhile 
mentioning that the above expression is applicable even when $R(\omega)$ and 
$M(\omega)$ have different normalizations.  The entropy measures how much the response 
function differs from the model.  It vanishes when $R(\omega)=M(\omega)$, and
is negative when $R(\omega) \neq M(\omega)$.  The maximum entropy method
adds to the simple $\chi^2$ minimization the use of the prior information that the
response function can be interpreted as a probability distribution function.
We employ {\it historic maximum  entropy} by minimizing $\alpha\, S-\chi^2/2$ with
the parameter $\alpha$ adjusted to make the $\chi^2$ equal to one.
While more refined methods relying on Bayes statistical inference
have been developed, we found historic maximum entropy to be simple to implement
and adequate for our purposes.

As a first case we consider the electromagnetic response of $^4$He.
We generated a set of $N_E \simeq  2500$ GFMC estimates of the
Euclidean response functions, obtained from independent imaginary-time
propagations on a grid of $\tau$ points uniformly distributed between 0 to
0.05 MeV$^{-1}$ with $\Delta \tau=0.0005$ MeV$^{-1}$.  Let $E^{(n)}_i = E^{(n)}(\tau_i)$
be the Euclidean response function corresponding to the $n^{\rm th}$ GFMC propagation.
The average Euclidean response function and covariance matrix elements are given by 
\begin{align}
\overline{E}_i&=\frac{1}{N_E} \sum_{n=1}^{N_E} E^{(n)}_i\ ,  \\
C_{ij}&=\frac{1}{N_E(N_E-1)}\sum_{n=1}^{N_E} E^{(n)}\!\!\left[ \overline{E}_i-E_{i}^{(n)}\right]\!\!
\left[\overline{E}_j-E_{j}^{(n)}\right]\, .
\end{align}
In general, the covariance matrix is non-diagonal because of correlations
between different $\tau_i$, and the full expression for the $\chi^2$ in Eq.~(\ref{eq:chi2})
has been used. 

\begin{center}
\begin{figure}[bth]
\includegraphics[width=8.5cm]{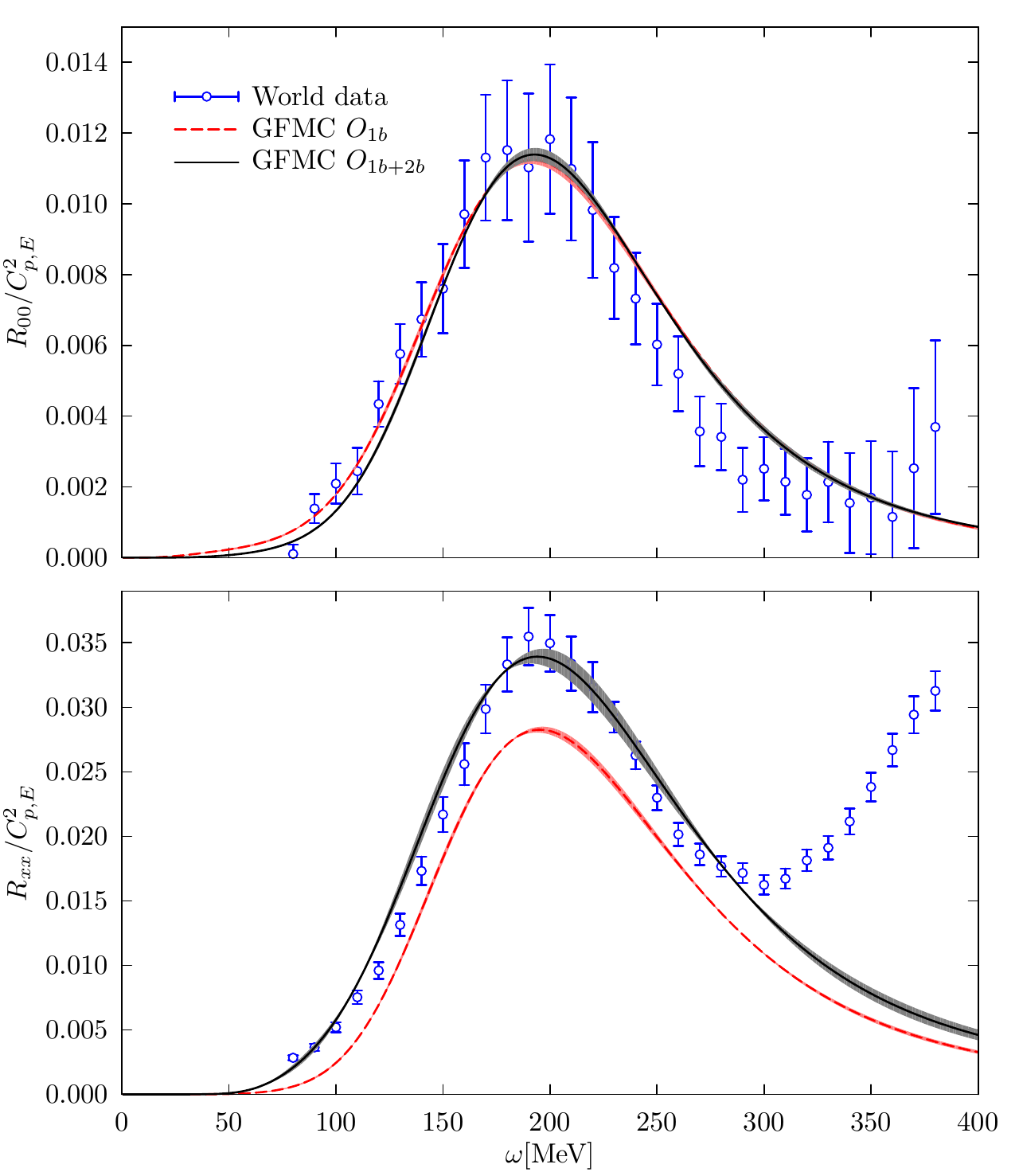}
\caption{(Color online) Electromagnetic longitudinal (top panel) and transverse (lower panel)
response functions of $^4$He at  $q=600$ MeV.  Experimental data are from Ref.~\cite{Carlson:2002}.}
\label{fig:f5}
\end{figure}
\end{center}

The $^4$He longitudinal and transverse response functions (at $q=600$ MeV),
obtained from inversion of $E_L(q,\tau)$ and $E_T(q,\tau)$, are shown in
Fig.~\ref{fig:f5}. The inversions are, to a very large degree, insensitive
to the choice of default model response~\cite{Lovato:2015}. Results obtained with one-body only (dashed line)
and (one+two)-body (solid line) currents are compared with experimental world data~\cite{Carlson:2002}
(empty circles).  There is excellent agreement between the full theory and experiment.
Two-body currents significantly enhance the transverse response function, not only in
the dip region, but also in the quasi-elastic peak and threshold regions, providing
the missing strength needed to reproduce the experimental results.
The band in Fig.~\ref{fig:f5} provides an estimate for the dependence
of the full results on the adopted default model, either
a flat $M(\omega)$ or a gaussian one~\cite{Lovato:2015}.  The model
dependence is quite small.

On the
basis of the present $^4$He and $^{12}$C calculations, a consistent picture
of the electroweak response of nuclei emerges, in which two-body terms in
the nuclear electroweak current are seen to produce significant excess transverse
strength from threshold to the dip region and beyond.  Such a picture is at
variance with the conventional one of inclusive quasi-elastic scattering, in which
single-nucleon knockout is expected to be the dominant process in this regime.
With the exception of the leading relativistic corrections contained in the nuclear electroweak
currents (see Ref.~\cite{Shen:2012}), the present calculations are based on a nonrelativistic
approach.  Naive kinematical considerations would lead one to expect the quasi-elastic
peak position in Fig.~\ref{fig:f5} to be at $q^2/(2\,m)+\Delta E\simeq 211$ MeV for $q=600$ MeV---we
take $\Delta E \simeq 20$ MeV to be the separation energy of $^4$He into a 3+1
cluster.  The calculated response functions appear to peak at lower $\omega$, in
fact close to $\omega_{\rm qe}+\Delta E \simeq 195$ MeV. The width
of the quasi-elastic peak is also seen to be correctly reproduced---the nonrelativistic
Fermi gas fails to predict this quantity at momentum transfers $q\sim 600$ MeV 
as in Fig.~\ref{fig:f5}.
Thus,
even at these relatively high momentum and energy transfers,
the nonrelativistic dynamical framework adopted here may be more robust than
comparisons between nonrelativistic and relativistic Fermi gas models would lead
one to conclude~\cite{DePace:2003}.

A direct evaluation of the $^{12}$C response functions via these 
same methods would require about 100 million core hours. We are examining
improved methods including the use of correlated sampling that could
improve the efficiency of this inversion. We are also exploring methods
to extend these results to larger nuclei.
\acknowledgments
We thank I. Sick for providing us with the data on the 
response functions of $^{12}$C and $^4$He. Useful discussions 
with O.\ Benhar, J.\ Gubernatis, and N.\ Rocco are also gratefully 
acknowledged.
This research is supported by the U.S.~Department of Energy, Office of Science, Office of
Nuclear Physics, under contracts DE-AC02-06CH11357 (A.L. and S.C.P.),
DE-AC02-05CH11231 (S.G.~and J.C.), DE-AC05-06OR23177 (R.S.), the
NUCLEI SciDAC program and by the LANL LDRD program.
Under an award of computer time provided by the INCITE program,
this research used resources of the Argonne Leadership Computing
Facility at Argonne National Laboratory, which is supported by the
Office of Science of the U.S. Department of Energy under contract
DE-AC02-06CH11357.  We also used resources 
provided by Los Alamos Open Supercomputing, and by Argonne's LCRC.
This research used resources of the National Energy Research
Scientific Computing Center, which is supported by the Office of
Science of the U.S. Department of Energy under Contract No. 
DE-AC02-05CH11231.

\bibliography{biblio}

\end{document}